\documentclass[twocolumn,showpacs,preprintnumbers,amsmath,amssymb]{revtex4}

\usepackage{graphicx}% Include figure files
\usepackage{dcolumn}% Align table columns on decimal point
\usepackage{bm}% bold math

\begin{document}

\title{On the abuse and use of relativistic mass}
\author{Gary Oas}
\affiliation{%
Education Program for Gifted Youth \\  Stanford University}
 \email{oas@epgy.stanford.edu}

\date{\today}

\begin{abstract}
The concept of velocity dependent mass, relativistic mass, is examined and is found to be inconsistent
with the geometrical formulation of special relativity.
This is not a novel result; however, many continue to use this concept and some have even attempted to establish it as the basis for special relativity. 
It is argued that the oft-held view that formulations of relativity with and without relativistic mass
are equivalent is incorrect. 
Left as a heuristic device a preliminary study of first time learners suggest that misconceptions
can develop when the concept is introduced without basis.
In order to gauge the extent and nature of the use of relativistic mass a survey of the literature
on relativity has been undertaken. The varied and at times self-contradicting use of this concept
points to the lack of clear consensus on the formulation of relativity.
As geometry lies at the heart of all modern representations of relativity, it is urged, once again, that the use of the concept at all levels be abandoned. 
\end{abstract}


\maketitle

\section{Introduction}  

The year 2005 has been deemed the ``World Year of Physics'' in conjunction with the 100th anniversary of Einstein's ``Annus Mirabilis.'' As a result, numerous discussions of Einstein's achievements will ensue throughout the year. There will, no doubt, be numerous presentations of the theory of relativity and quantum mechanics in newsprint, books, and on television. It is the responsibility of physicists to educate the lay-public about these ideas and their importance to society. However, it behooves the physics community to insure that such presentations are free of inaccuracies and hyperbole. 
Action must be taken when broad, inaccurate assertions are put forth
(such as stating everything is random in quantum mechanics, or that Einstein did not believe in quantum mechanics\cite{pais}).  

There is one concept that has been ingrained into the collective mindset of not only lay-people but also many working physicists. This is the notion of relativistic mass; a moving object's mass increases with velocity with respect to an observer considered to be at rest, 
\begin{equation}
		m(v) = {m_0\over\sqrt{(1-v^2/c^2)}}.   \label{rm}
\end{equation}

The decision whether or not to introduce relativistic mass in pedagogical expositions is still not universally agreed upon. 
The debate on the usage of this concept has been going on since the early days of relativity. On both sides of the issue there
have been advocates insisting their view is the correct one. Some have relied on historical precedent as a rationale for its
inclusion.

\begin{quote}
{ \small Whether or not to speak of velocity-dependent mass is largely a matter of taste. Although it is currently unfashionable to do so, Einstein did and we shall as well.} (pg 212 of {\em Understanding Relativity}\cite{sartori}){\small (Q1)}.
\end{quote}

Reliance upon such a controversial history for validation 
is not only pedagogically suspect but is, in this instance, incorrect; Einstein was {\it not} in favor of using relativistic mass, except in his earliest works. 
In fact, one would be hard-pressed to find an explicit statement of relativistic mass in Einstein's work after his 1906 paper on transverse and longitudinal mass\cite{einstein}. Later in life, he addressed the issue directly in a letter to L. Barnett.

\begin{quotation}
	{ \small It is not good to introduce the concept of the mass $M = m/(1-v^2/c^2)^{1/2}$ of a moving body for which no clear definition can be given. It is better to introduce no other mass concept than the 'rest mass' m. Instead of introducing M it is better to mention the expression for the momentum and energy of a body in motion.}  (quote from \cite{okun})(Q2) 
\end{quotation}

In the late 1980's a renewed effort was made to dissuade the physics community from introducing this concept\cite{okun}\cite{okun2}\cite{adler}\cite{taylor2}. These attempts spawned a vigorous debate on the interpretation and usefulness of relativistic mass. In section \ref{litsearch} evidence will show that this debate has
had some influence in diminishing the use of this concept in textbooks; however, 
it is clear that these efforts have gone unheeded by many. 

It is not the intent here to revisit all of the previous arguments but to expand upon them in order to convince the need and urgency of the issue. (See Lev Okun's extensive arguments\cite{okun2}
for a thorough discussion).
 The need for continued discussion stems from recent work attempting to legitimize the concept as a basis for special relativity, and also the continual flood of books utilizing the concept geared for the general public.

\section{What are the opposing viewpoints?}

Most of the previous arguments debating the use of relativistic mass have oversimplified the difference in viewpoints in question; it is not merely a matter of taste whether to employ the concept or not.
Thus, the first task at hand is to define the opposing viewpoints. 

\begin{description}
\item[Non-relativistic mass view, Geometric Formulation (G):]
Central to the geometric formulation (G) is that the isotropy of space and uniformity of space and time, along with the 
notion of causuality\cite{reichenbach}\cite{hawking2}\cite{bombelli}, is sufficient to lead to a four-dimensional spacetime endowed with a Lorentzian signature metric. 
The principle of relativity yields a consistent kinematics incorporating the effects of time dilation, length contraction, addition of velocities, etc. 
The Lorentz transformation is seen as a direct consequence of the hyperbolic geometry.
The extension to dynamics is immediate by postulating the primitive concepts of mass and momentum. In order to have a Lorentz covariant 4-momentum, the mass is found to be a Lorentz invariant. The validity of this approach will be assumed in what follows.

\item[Relativistic Mass (RM):]
Here we define the concept of relativistic mass (RM) as the relation (\ref{rm}). This is a dynamical concept without specification to kinematics. Exactly how RM is invoked in relation to kinematics will be explored below, as this is the purpose of this paper. 
	
\end{description}

In regards to the antiquated concepts of longitudinal and transverse mass, they shall not be considered here. C. Adler\cite{adler} demonstrated how these must be accepted if one is to propose an inertial relativistic mass.

\subsection{How is RM invoked?}
 
	In most introductory treatments of special relativity the exact nature of how RM is arrived at is varied and often vague. Thus, the task at hand is to explore how this concept is to be incorporated within the theoretical structure of special relativity. There are two possibilities,
\begin{enumerate}
\item RM supplants the geometric formulation as a primitive concept, 
\item RM is introduced in conjunction with the geometric formulation. 
\end{enumerate}

However one may feel about the validity of either possibility, both are taken seriously in different works
and thus it is important to examine each closely. We begin with the foremost reason that RM is introduced in works directed at the general public; to explain why no material object can travel at or beyond the speed of light (this concept will be denoted as Noc). 
Given that G implies time dilation, which yields Noc, the question arises as to exactly how RM is invoked to give Noc. 
There are two possibilities:
either RM itself is the causation of Noc (1. above), or RM is considered in conjunction with the proper kinematics to yield Noc (2. above). If neither position can be made tenable all that is left is a heuristic device.

\subsubsection{Can RM supplant G?}

Those adherents of the first position might be inclined to state that there are two different, yet equivalent, viewpoints of the same phenomena. In regards to Noc, a common argument is provided in the following quote.

\begin{quote}
{\small ...Einstein's equation gives us the most concrete explanation for the central fact that nothing can travel faster than light speed. You may have wondered, for instance, why we can't take some object, a muon say, that an accelerator has boosted up to 667 million miles per hour -- 99.5 percent of light speed -- and ``push it a bit harder,'' getting it to 99.9 percent of light speed, and then ``{\it really} push it harder'' impelling it to cross the light-speed barrier. Einstein's formula explains why such efforts will never succeed. [\ldots] But the more massive an object is, the harder it is to increase its speed. Pushing a child on a bicycle is one thing, pushing a Mack truck is quite another. So, as a muon moves more quickly it gets ever more difficult to further increase its speed.}  (pg 52 of {\em The Elegant Universe}\cite{greene1})(Q3)
\end{quote}

Some suggest that this viewpoint obviates the need to introduce time dilation, Lorentz transformations, or other seemingly complicated concepts in demonstrating Noc\cite{sandin}, that is, RM is a sufficient to explain Noc. 
The problem with this explanation is that, by itself, relativistic mass does not prevent the observation of superluminal objects. For it does not imply the addition of velocity formula and therefore two observers in two IRFs moving in opposite directions may observe the other to be traveling at a speed greater than c.  Thus it is not a sufficient reason for the existence of a fundamental speed limit for any object. Of course along with the principle of special relativity, and its implication that the speed of light is an invariant, relativistic mass does allow for Noc. However the constancy of the speed of light is sufficient, in itself, to give Noc\cite{mermin}. To repeat, RM in itself is not sufficient to give Noc, yet if it is introduced along with the principle of special relativity it is superfluous.

Throughout the past century there have been a few attempts to derive special relativity from
classical mechanics and the relativistic mass relation, (\ref{rm}), alone\cite{terletskii}\cite{jammer}\cite{landau}.
	
\begin{quote}
{\small For those who want to learn just enough about it so they can solve problems, that is all there is to the theory of relativity -- it just changes Newton's laws by introducing a correction factor to the mass.} \\
(pg 15-1 of {\it The Feynman Lectures on Physics}\cite{feynman1}) (Q4)
\end{quote}

No attempt has been entirely successful, however this does not prevent
some from trying. It is instructive to examine the latest such attempt to see how this program 
fails.

\noindent
{\bf Jammer.}
In his more recent book on concepts of mass\cite{jammer}, Max Jammer delves extensively into the numerous arguments related to relativistic mass. The conclusion reached is that both formalisms are valid and that the difference is ``ultimately the disparity between two competing views of the development of physical science.'' To this end he cites the work of B. V. Landau and S. Samanthapar\cite{landau} in which the addition of velocity relation and Lorentz transformation are said to be derived from Newtonian mechanics and the relativistic mass formula alone. As Jammer puts it ``The fact that the Lorentz transformation and relativistic mass equation mutually imply one another seems to indicate that the relation between these two is more intimate than commonly thought''. This result would formally show the validity of relativistic mass as a possible basis for special relativity, thus making the RM formalism and the geometric formalism complementary views of special relativity.

	A close examination of Landau and Samanthpar's derivation finds that the hyperbolic geometry of spacetime is the basis for their result and not the relativistic mass formula. 
By considering conservation of momentum in a simple two-particle process, the authors implicitly invoke the geometry of spacetime by redefining, without explanation, the velocities as hyperbolic tangents (i.e. ${u_1\over c} = tanh\alpha$, etc.).
Then by relying on hyperbolic trigonometric relations, hyperbolic spacetime has been introduced, automatically yielding the relation they desire\cite{taylor1}. Thus this derivation does not provide a basis for Jammer's claim since the Lorentz transformation arises from the geometry and not from the relativistic mass formula. 

	To support his claim further, Jammer derives the relativistic mass relation from the Lorentz transformation. However his derivation is, at the outset, set up to give a relativistic mass once the spatial components of the 4-momentum are defined. For all agree (modulo interpretations of $m_0$) that the Lorentz transformations yield a 4-momentum that must be of the form,
\begin{equation}
p^\mu  =  \gamma m_0 u^\mu,   {\label{4mom}}
\end{equation}
where $u^\mu = {dx^\mu\over dt}$. This derivation precludes the interpretation of a proper 4-velocity, $v^\mu = {dx^\mu\over d\tau}$, that stems from G and which results in (\ref{4mom}) without need for relativistic mass. Thus the derivation does not prove the claim that the Lorentz transformations unequivocally lead to a relativistic mass, nor does it lend credence to the idea that these two concepts are intimately related.

V. Petkov has just written a text on special relativity\cite{petkov} in which he endeavors to
give a consistent ontological foundation of spacetime. Though more philosophical in nature,
it is a serious exposition but does ascribe to a more radical RM viewpoint: 
``It should be stressed that the resistance arises {\it in} the particle; it does not come from the geometric properties of spacetime."
As this argument is different in nature to most others, and rather intricate, a detailed
exploration will need to be done at a later time.

Feynman's assertion above (Q4) is incorrect (for example $p^\mu_{rel}\neq m(v) v^\mu_{rel}$).
Clearly the notion that relativity can be derived from the relativistic mass relation and classical principles alone is untenable. Without the kinematical effects of time dilation one is forced to accept superluminal, imaginary mass objects
that could easily be generated by a suitable choice of reference frames. All one need do to observe such objects in such a world is to walk briskly in a direction 
opposite that of electrons near the end of a linear accelerator such as SLAC.
It is the geometry of spacetime and not RM that limit material objects to less than the speed of light.

\subsubsection{Can RM be consistent with G?}

With the failure to base special relativity upon relativistic mass, independent of G,
one is lead to consider the second possibility of invoking RM, as a primitive concept, in conjunction with G. 
\begin{quote}
{\small And it's not just distance and time that change. Special relativity also shows that as an object travels faster, its mass increases. But as the mass of an object increases, it takes more and more energy to increase its speed any further. Eventually, as the object gets close to the speed of light, it becomes so massive that no amount of energy will make it go any faster. This means that the speed of light is a universal speed limit which nothing with mass can break.}
({\em Einstein Year UK} web site\cite{einsteinwebsite}\label{q5})(Q5)
\end{quote}

To see how relativistic mass is arrived at in more formal texts, we need to understand the
main motivation behind its introduction --maintaining familiar, Newtonian-like, expressions for velocity and momentum.

\begin{quotation}
{\small \ldots we shall show that it is possible to preserve the {\it form} of the classical definition of the momentum of a particle, ${\bf p} = m{\bf u}$, where ${\bf p}$ is the momentum, $m$ is the mass, and ${\bf u}$ the velocity of a particle, and also to preserve the classical law of conservation of momentum of a system of interacting particles, provided that we modify the classical concept of mass.}\cite{resnick}(Q6).
\end{quotation}

	This predilection towards Newtonian momentum is back of most derivations of relativistic mass. The reason stems from the early history of special relativity, especially the efforts of Lewis and Tolman\cite{lewis}\cite{tolman}\cite{tolman2}, that were picked up by later authors\cite{feynman1}\cite{french}\cite{resnick}. Just as in Jammer's case above, these derivations posit a Newtonian form for the 3-momentum, 
\begin{equation}
p_x = mu_x,			\label{Rmmom}
\end{equation}
and then impose conservation of momentum.
Naturally the result found is that the new relativistic expression for momentum, that is Lorentz covariant, must be of the form (\ref{4mom}). From his derivation, Tolman (1912, as cited in Jammer\cite{jammer}) makes the bold assertion that ``the expression $m_0(1-v^2/c^2)^{-1/2}$ is best suited for THE mass of a moving body''. 

\vspace{10 pt}\noindent
{\bf Breaks Lorentz Covariance.}
By asserting (\ref{Rmmom}) as the definition of relativistic momentum, one is forced to adopt a primitive concept of an improper 4-velocity; for $u^\mu = \frac{dx^\mu}{dt}$ is not Lorentz covariant. It is easy to see that Lorentz transforming from a frame where an object is moving with a 3-velocity $u$ to its rest frame that,
\begin{equation}
u'^\mu = \Lambda^\mu_\nu (u)u^\nu = ({c\over \gamma},0,0,0), 
\end{equation}
whereas in its rest frame it is required to be $(c,0,0,0)$. 

 The improper velocity being a direct result of the imposition of RM means that RM is at odds with the accepted kinematics of special relativity.
The notion that the choice whether to utilize relativistic mass or not is merely a matter of taste is seen to be fallacious: The true nature of special relativity stems from geometry (or kinematics).

\begin{quote}
{\small Indeed, it should be noted that, whether we identify the factor $1/\sqrt{1-\beta^2}$ with mass or with velocity, the origin of this factor in collision measurements is kinematical; that is, it is caused by the relativity of time measurements.} (pg 199 of \cite{resnick})(Q7).
\end{quote}

The above method of breaking Lorentz covariance by hijacking the dilation factor away from the proper 4-velocity and assigning it, for reasons of familiarity or heuristics, to the mass is exemplified in Wolfgang Rindler's classic text\cite{rindler}. This is an excellent formal introduction to relativity that delves into the geometric formalism with great clarity and depth. However, imbedded within this discussion are continual dichotomous references to relativistic mass. 
\begin{quotation}
{\small We begin by assuming what we already know, that associated with each particle there is an intrinsic positive scalar, $m_0$, [...] This allows us to define the 4-momentum $P$ of a particle in analogy to its 3-momentum, 

  $ P = m_0U$ 
  
    $ U$ being the 4-velocity.} 
\end{quotation}
One paragraph later it is amended,
\begin{quotation}
{\small we find the following components for P:
$P = m_0U = m_0\gamma(u)(u,c) =: (p,mc)$. } (Q8).
\end{quotation}

This relativistic shell game would not be too objectionable if the improper 4-velocity were not to be considered. However, the geometric formulation has been usurped by the introduction of an improper velocity that does not transform under a Lorentz transformation. 

If this concept is so central to relativity it ought to be employed throughout any discussion of relativity. However it is nearly always abandoned soon after its introduction.

\begin{quote}
{\small  In section 2.12 we assumed a relativistic momentum of the form $p = m(v)v$ and used this to look for an expression for the relativistic mass $m(v)$ such that the conservation laws for mass and energy would hold in an interaction. [...] The expression for momentum is simply rest mass multiplied by $\gamma v$ which we now recognise [sic] as the first three components of the 4-velocity.} (Pg 161 of \cite{adams})(Q9).
\end{quote}
\begin{quote}
	{\small Some authors use the relativistic mass but we shall do so only in the next two sections.} 
(pg 49 of \cite{williams})(Q10).
\end{quote}

	Clearly those who utilize the concept do not hold it dear. The concept has only limited explanatory power and is inconsistent with the geometrical formulation. 
If one can provide simple alternative explanations for Noc, as well as other aspects of relativity,
while reinforcing the spacetime concept, why introduce RM in the first place?

\section{Pedagogical arguments: Can RM be a Useful Heuristic?}
 
	Having found that a consistent theory of special relativity that is based upon proper relativistic quantities and relativistic mass can not be formulated, one is left to wonder why it is still considered with such passion.  If it can not be considered as a primary concept of the theory, all that is left is to consider it as a heuristic. This is troublesome for it introduces a view that is at odds with the formal theory. 
An analogy might be to insist 
on discussing centrifugal forces, without reference to non-inertial frames, for ``it appears'' as if there is a force in a non-inertial frame, however few physicists would find this an advisable approach. However this does not deter all from abandoning this approach. In his response to  Okun's call to abandon the concept, Rindler\cite{rindler2} offered the following,

\begin{quote}
{\small To me, $m = \gamma m_0$ is a useful heuristic concept. It gives me a feeling for the magnitude of momentum ${\bf p} = m{\bf v}$ at various speeds [\ldots] I will confess to even occasionally using the heuristic concepts of longitudinal mass $\gamma^3m_0$ and transverse mass $\gamma m_0$ to predict how a particle will move in a given field of force.} (Q11)
\end{quote}

\vspace{10 pt}\noindent	{\bf Potential misconceptions.}
The concept of relativistic mass is often thrust upon the reader with little or no justification 
(a typical example is the popular {\em The Physics of Star Trek}\cite{krauss}). Not only does this force the reader to accept it on faith, it leaves a seed of misconception that in the object's rest frame the mass is increasing. Consider the following passages,

\begin{quote}
{\small Because of the equivalence of energy and mass, the energy which an object has due to its motion will add to its mass. In other words, it will make it harder to increase its speed. [\ldots]
As an object approaches the speed of light, its mass rises ever more quickly, so it takes more and more energy to speed it up further. It can in fact never reach the speed of light, because by then its mass would have become infinite, and by the equivalence of mass and energy, it would have taken an infinite amount of energy to get it there. For this reason, any normal object is forever confined by relativity to move at speeds slower than the speed of light.}
\vspace{-10 pt} \begin{flushright} (pg 20-21 of {\em A Brief History of Time} \cite{hawking})(Q12)\label{q12}\end{flushright}
\end{quote}

\begin{quote}
{\small The faster something moves the more energy it has and from Einstein's formula we see that the more energy something has the more massive it becomes. Muons traveling at 99.9 percent of light speed, for example, weigh a lot more than their stationary cousins. In fact, they are about 22 times as heavy--literally.}
\vspace{-10 pt} \begin{flushright} (pg 52 of {\em The Elegant Universe} \cite{greene1})(Q13).\end{flushright}
\end{quote}

\begin{quote}
{\small Special relativity - how clocks can run slow, and objects can shrink and gain mass at the same time}
\vspace{-10 pt}\begin{flushright} (UK Einstein Year Web Site, \cite{einsteinwebsite} )(Q14)\end{flushright}
\end{quote}

The implication within these passages is that the object is {\it literally} growing more massive. These statements are indeterminate for the reference frame is not stated. This is not true of all works\cite{falk}, however this form of relativistic indetermination pervades the literature. 

For the uninitiated this situation may be hard to grasp --exactly how is it becoming more massive? Are there more atoms in the object? How does one actually measure this increase? Can one actually measure it? Without further explanation one is left with a physical picture that is vastly different than what the author intended. Or is it? To see that RM leads to misconceptions, not only
among students but authors as well, all one need do is purview some popularizations of relativity. Consider two examples that highlight this problem.

\begin{itemize}
\item {\em Relatively Speaking} by Eric Chaisson\cite{chaisson} 

On page 62 of this pedestrian introduction to relativity there is a series of three illustrations of a girl standing on a scale. The intent is to convey observations when she is at rest and travelling at two different speeds  (labeled ``fast'' and ``faster''). The girl is seen to Lorentz contract as expected and the scale indicates her weight to increase -- the needle points to successively higher values. This is exactly the type of misconception that is reinforced by this concept. If this image is to represent the girl in various IRFs then the needle pointing to different weights violates the primal nature of events, one of the most basic tenets of relativity.
 
\item {\em Space and Time in the Modern Universe} by P.C.W. Davies\cite{davies}

On page 45 there is depicted a stick figure, holding a string with a ball on the end, 
standing on a scale indicating 1 kg. Adjacent to this image is an image in which the same person
is twirling the ball over its head and the scale now indicates 5 kg -- supposedly reflecting the
increase with mass with velocity. The manner in which the ball was set into motion is not stated.
If an outside agent set it in motion then there is no controversy that the mass of the isolated person-ball system has increased. The RM viewpoint is that the energy input went into the kinetic energy
of the ball plus its mass increase while the G viewpoint is that the energy went to the kinetic energy alone.
If the person sets the ball in motion (which seems to be the implication), then there is no
valid interpretation of the depiction, for energy (and hence mass) is conserved for this isolated system. 

\end{itemize}

This last example leads to a subtle point about the definition of mass and the definition of object (or system of objects) and is stated by some\cite{sandin}\cite{whitaker} to be a pitfall of the non-RM viewpoint. However, the definition of the mass of a system as the proper energy divided by $c^2$ 
--regardless of the motion of the system's center of mass -- is clearly described by Einstein\cite{einstein2}.
The energy of an object as measured in other frames of reference is not inherent to the
object itself (is not proper).
There are no conceptual inconsistencies with this approach (see Harris\cite{harris} for
a clear textbook treatment).

The two viewpoints of kinetic energy lead to a differing ontology of spacetime. 
\begin{eqnarray}
K_{RM} & = &(1-1/\gamma )m_{rel}c^2, \\
K_{G} & = &(\gamma - 1)mc^2.
\end{eqnarray}
The divergence of kinetic energy as light speed is approached has its cause in the diverging mass in the RM viewpoint and the geometry of spacetime in the G viewpoint. This, once again, points to the incompatibility of the RM viewpoint with a geometrical view of spacetime.

\subsection{Preliminary Research Into Misconceptions}

It has been argued that the concept of relativistic mass can not be a part of any complete, consistent, formulation of relativity. 
Two examples of popular introductions to relativity demonstrated that the use of relativistic mass may lead to misconceptions, even among experts. In order to fully gauge the potential for such misconceptions to arise, more research is needed. If it can be clearly demonstrated that RM leads to 
problematic thinking, regardless of any purported pedagogical benefits, then one must seriously examine its continued use.

A preliminary investigation into misconceptions developing among first time learners of relativity has been undertaken,
consisting of a brief survey given to 164 students\cite{survey}.
Students were randomly given one of three short passages
from popularizations of relativity geared for general audiences; consisting of quotes (Q12)
on page \pageref{q12}, (Q5) on page \pageref{q5}, and the following, (Q15), taken from a
recent popularization of modern physics.

\begin{quote}
{\small Einstein also showed that the mass of an object moving at close to the speed of light, as seen by an outside observer, increases. [...]
It explains, among other things, why the speed of light serves as the ultimate speed limit in the universe. Suppose you're in a spaceship, approaching the speed of light. You think `I'll just step on the accelerator a little harder and I'll pass that pesky speed limit, no problem.' But it won't work: to make your craft move faster, you have to use energy; the more massive the spaceship, the more energy you need. And, thanks to Einstein's special relativity, the mass keeps increasing, so you need more and more energy to further boost the speed. And you'll never quite make it. If you reached the speed of light, your spaceship, as seen by an outside observer, would have an infinite mass and it would have taken you an infinite amount of energy to get there.}
(from {\em Universe on a T-Shirt}\cite{falk}) (Q15)
\end{quote} 

After reading their brief passage the students were queried about two identical twins, one of which (Al) is aboard a rapidly moving train car while the other (Bob) watches from the ground. Given that Bob sees his own scale
to read 200 lbs, the students were asked; a) As Bob looks at Al's scale what does he see it to indicate?; b) As Al looks at his own scale what does he see it to indicate? The choice of response consisted
of less than, greater than, or equal to, 200 lbs. The results are given in the following table.

\begin{table}[h]
\begin{center}  \label{table1}
\begin{tabular}{|r||c|c|c|}
\hline
 &\multicolumn{3}{c|}{Response to b):} \\ 
Response to a): & $>$ 200 lbs. & = 200 lbs. & $<$ 200 lbs. \\ \hline\hline
 $>$ 200 lbs. & $24_{ (9,12,3)}$ & $70_{(14,30,26)}$ & $8_{(4,2,2)}$\\ \hline
 = 200 lbs. & $11_{(4,5,2)}$ & $28_{(13,6,9)}$ & $7_{(3,1,3)}$\\ \hline
$<$ 200 lbs. & $2_{(0,2,0)}$ & $9_{(3,3,3)}$ & $5_{(0,1,4)}$ \\  \hline
\end{tabular} 
\end{center}
\caption {Results of survey of 164 students. Rows refer to answers to question (a) -- the scale reading of Al as viewed by Bob, and columns correspond to (b) --the scale reading of Al as viewed by Al himself. 
The total number of responses is given and
the subscript triplet gives the breakdown by passage given ((Q5), (Q12), (Q15)).}
\end{table}

	Of note is that less than a fifth of the students gave the completely correct response, (=,=),
and that more than a third indicated that the weight changes in the rest frame Al. 
Also interesting is that (Q15) is the only quote of the three to explicitly state the
relativity of the observations in different frames, yet less than a fifth (9/52) gave the correct
answer.

While these are only preliminary results, and do not warrant a firm conclusion, they do suggest  
that invoking the concept of relativistic mass may create more confusion among the lay reader. 
Clearly more thorough research is needed and is currently underway.

One might argue that the misconception discovered here is no different than what is often
found in regards to time dilation or length contraction -- a strong bond to Galilean relativity 
that is difficult to overcome. 
The effects of time dilation and length contraction can be explictly measured in simple ways.
Any change in mass can not be measured directly but only inferred.
In addition, the use of relativistic mass is based on
heuristic grounds and is not a fundamental part of the theory. Why employ an unnecessary
concept that is furthering confusion?

\section{Survey of Literature}  \label{litsearch}

	The discussion up to this point has been intended to demonstrate that the concept of relativistic mass is problematic. For those that already understand this point we would like to point out that this notion is still prevalent in the literature on relativity. To this end the results of an extensive, yet not exhaustive, literature search are now discussed. 

	In total, 637 works were reviewed to determine whether the concept of relativistic mass is introduced. Those works that introduce RM but then discuss its shortfalls, or dissuade its use, are listed as not having introduced the notion. The primary concern is to identify those works that deem relativistic mass as a valid aspect of relativity. Emphasis has been placed on latest, available editions of works.
Due to its length, the reference list for these works, along with further commentary, is given in a separate report\cite{oas2}.

Of all the works examined, 477 relied upon the concept. This is not a very informative
fact in itself; it is more revealing to examine the historical trend among different types of works. 
To this end, the works were categorized into four broad
categories: texts devoted to special and/or general relativity, introductory and modern physics
textbooks, popularizations of relativity and physics, and a miscellany of other  works. This latter category
includes: philosophical, historical, religious, and science fiction books as well as textbooks that fall outside of those categories above (such as advanced physics, math and engineering). Articles and peer-reviewed journals were
excluded from this survey. One reason was simply to keep the task manageable, the other
was to limit to sources that are accessible to a wide population, not primarily scientists.

\begin{figure*}[htbp] %  figure placement: here, top, bottom, or page
   \centering
    \includegraphics[]{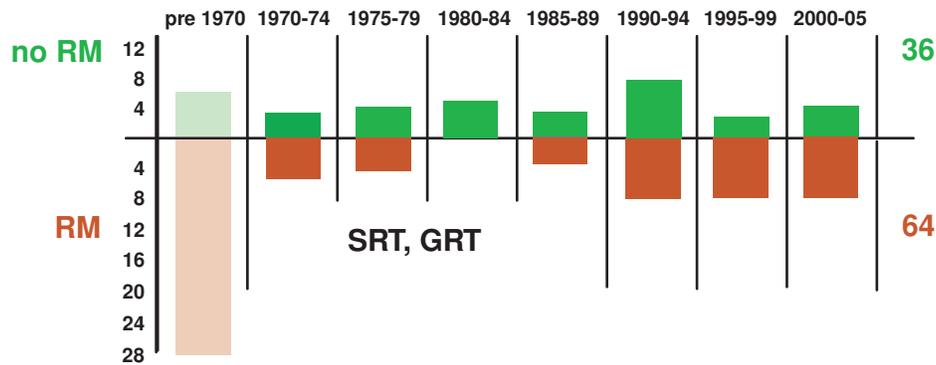} 
   \caption{Special and General Relativity Textbooks: 100 textbooks devoted to relativity displayed by year of publication and
classified as whether having introduced the concept of relativistic mass (RM) or not (no RM).}
   \label{fig1}
\end{figure*}

\begin{figure*}[htbp] %  figure placement: here, top, bottom, or page
   \centering
    \includegraphics[]{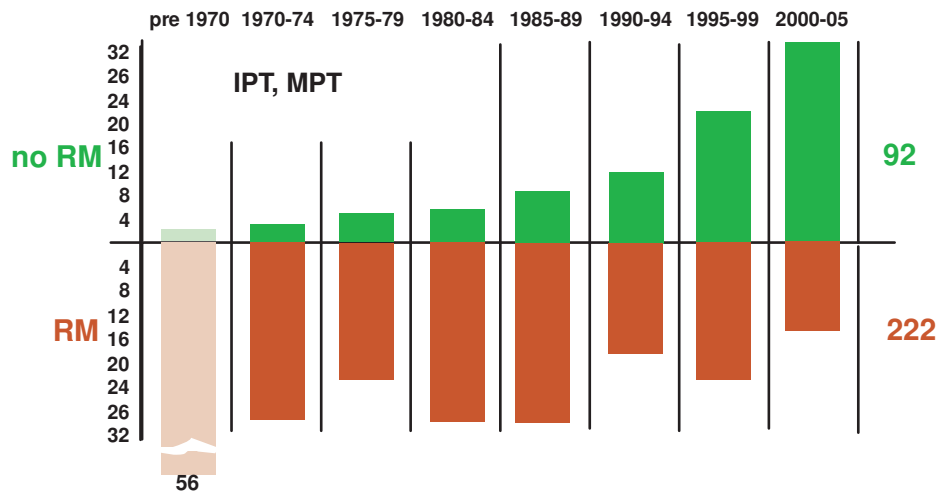} 
   \caption{Introductory and Modern Physics Textbooks:  314 editions of introductory and modern physics textbooks displayed by
year of publication of edition. }
   \label{fig2}
\end{figure*}

\begin{figure*}[htbp] %  figure placement: here, top, bottom, or page
   \centering
  \includegraphics[]{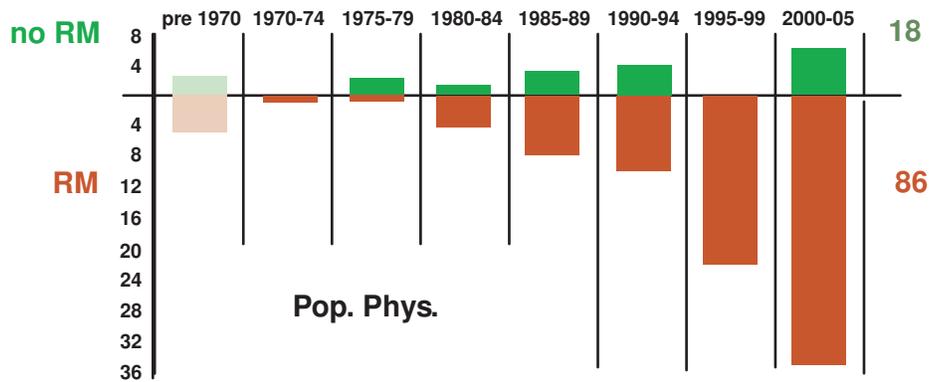} 
   \caption{Popularizations of Relativity and Physics: 103 books on science that are geared for the general public, displayed by year of publication.}
   \label{fig3}
\end{figure*}

\begin{figure*}[htbp] %  figure placement: here, top, bottom, or page
   \centering
\includegraphics[]{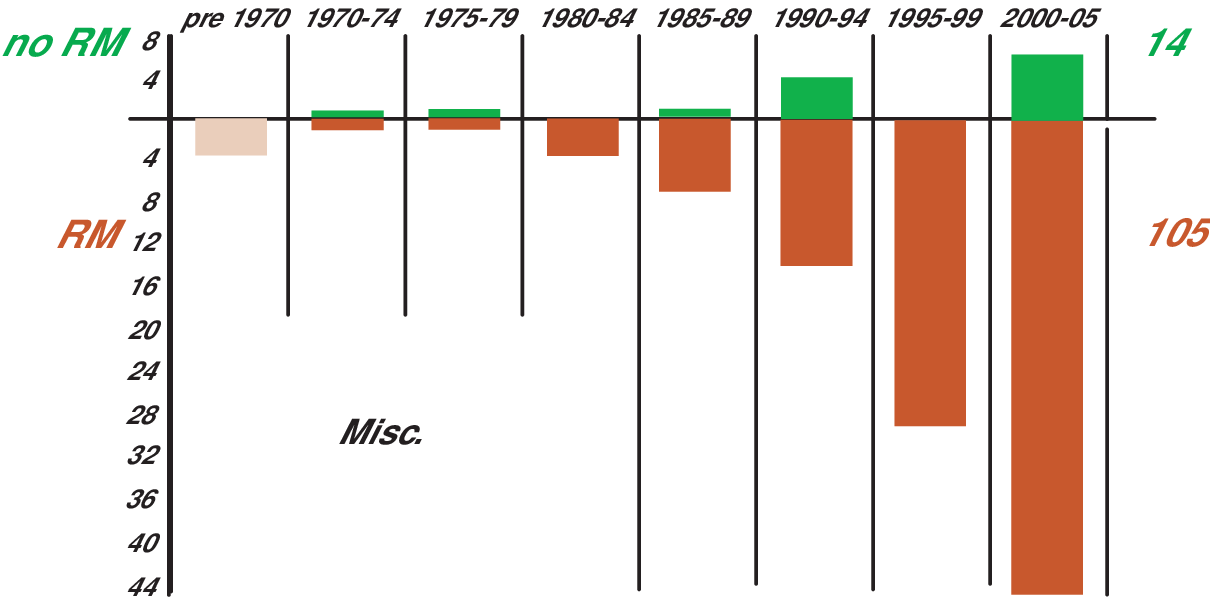} 
   \caption{Miscellaneous: 119 books on various topics such as: advanced physics, math texts, philosophical works, engineering texts, science fiction, religious, etc.}
   \label{fig4}
\end{figure*}

The historical trend of the use of relativistic mass is displayed in figures \ref{fig1} - \ref{fig4}. The
works were binned into five year increments beginning from 1970.

%Convert these images to eps

The last figure, \ref{fig4}, would appear to allude to a large bias in the treatment of the concept; however, 
one most be careful in drawing too strong of a conclusion from this portion of the survey. 
A large portion of the references were obtained from a search for relativistic mass. One should
not conclude from this set that the ratio of works in this category that use the concept to 
those that do not is growing rapidly. A valid conclusion is that the number of works in this category
that use this concept is growing. 

Certain trends are clearly indicated; the use of relativistic mass in introductory physics textbooks
has clearly been diminishing while its occurrence in more informal works is growing. A slightly 
more troubling result is the lack of any trend in more advanced texts devoted to relativity.

\subsection{A Closer Look at Some Textbooks}
Of the more significant trends, in terms of pedagogy, is the use of RM in 
introductory and modern physics textbooks. In addition to the above plot, figure 2,
it is of interest to see how individual textbooks have approached this concept
as they evolved into their later editions.
 In figure 5 are displayed some of the more popular 
physics textbooks categorized by year of edition and use of RM. 
The introductory physics texts have been categorized by their level, A, B, and C,
as defined by the College Board\cite{ap}. Category A being conceptual introductions;
B, algebra-based; C, calculus-based introductions; and an additional category D, representing
modern physics textbooks.

\begin{turnpage}
\begin{figure*}[htbp] %  figure placement: here, top, bottom, or page
\centering  
\includegraphics[height=5.5in]{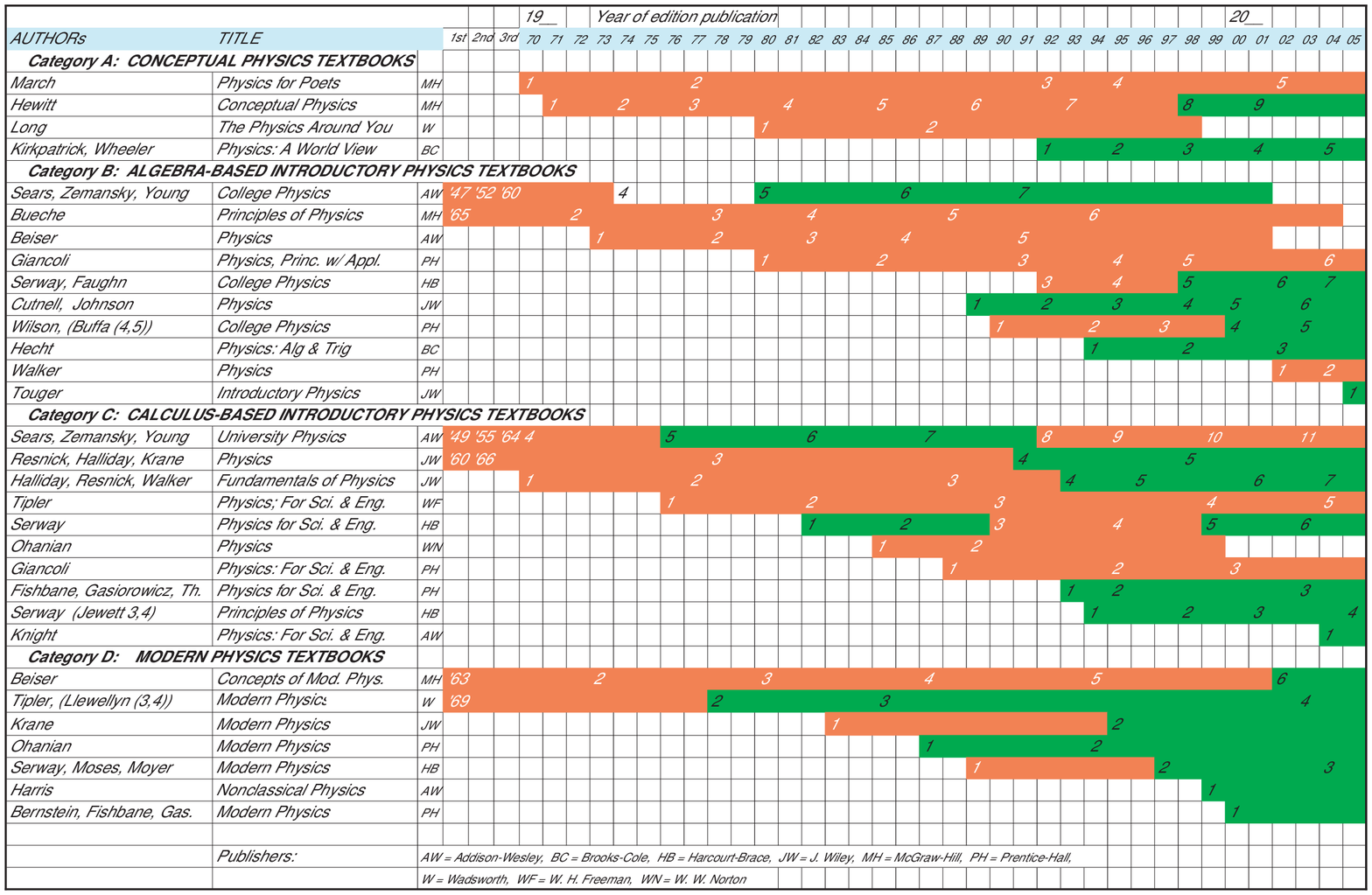} 
   \caption{ A selection of some popular introductory textbook editions categorized by year of publication and whether or not the concept of relativistic mass is used. Darker shading and white entires (orange online) indicates introduction of relativistic mass and lighter shading, dark letters (green online) indicates its absence. Omissions are left unshaded and last editions are given a shelf life of 10 years.}
   \label{fig5}
\end{figure*}
\end{turnpage}

In all of the hundred or so titles falling under this category only two have been found that moved
from a position of not employing the concept of relativistic mass to utilizing it.
The first is the popular {\em Physics: for Scientists and Engineers} by R. Serway\cite{serway} wherein the
third and fourth editions are the only ones to adopt the concept.
In fact, within these two editions, contradicting statements about the mass of an object are stated.
On page 1175 of the fourth edition  it is stated,
\begin{quote}
Finally, note that since mass $m$ of a particle is independent of its motion, $m$ must
have the same value in all reference frames.
\end{quote}
Two pages later it is claimed,
\begin{quote}
It follows that mass varies with speed (relative to the observer). We must therefore
distinguish between the {\bf rest mass,} $m_0$, which is the mass measured by
an observer at rest relative to the particle (and at the same location), and the mass
measured in real experiments.
\end{quote}
In the third edition the contradiction is more blatant; first, a mathematical expression for relativistic mass is given (page 1124) and then on page 1128 the first of the two quotes above appears.

The second case is the classic calculus-based textbook
 {\em University Physics} by Sears, Zemansky, and Young\cite{szy} 
 (carried over into later editions by the authors Young and Freedman). 
This series has enjoyed a very long history and has reversed its position on the concept twice, initially employing the
concept (like most texts of the time) and then moving away from it (5th ed.) only to 
return to its use in the latest editions. (This later inclusion can
be traced to the addition of T.R. Sandin as contributing author, an ardent promoter
of RM's use in pedagogical texts\cite{sandin}). These two texts aside, the overall trend has been
one of moving away from relativistic mass.

One of the purposes of this survey is to attempt to measure any effect of previous
calls to abandon this concept. Outside of the first edition
of {\em Spacetime Physics}\cite{taylor1}, most of the debate occurred between 1987 and 1992\cite{adler}\cite{okun}\cite{taylor2}\cite{rindler2}. The influence of these later calls
can be seen in the movement away from the concept in introductory and modern physics textbooks
beginning in the 1990s.
Thus it appears that such efforts were successful; however, among
formal textbooks devoted to relativity little change has occurred over the
past thirty years.
Worse yet, those works directed at the general public still, overwhelmingly, utilize this concept.
These diverging trends in the use of RM can be understood from the observation that much more attention is paid to physics education research when writing an introductory textbook as opposed to
an advanced text or popularization of physics. Reflection upon this research often lead to the concept's exclusion. 
More vocal calls for the abandonment of relativistic mass are clearly needed.
%An expanded, up-to-date, version is available on the web\cite{oas4}.

\subsection{A journey around the neighborhood}
A survey of texts over the past century does provide insight into the general viewpoint 
of relativistic mass among various writers. 
A different but poignant test is to ascertain what an average person, who desires
to learn about relativity, might encounter by accessing available resources.
To get a sense of such an encounter, I visited the local, chain, mega-bookstore
and surveyed the entire physics section. In this particular branch in downtown San Francisco,
 the section had an 
ample supply of books on physics, 351 in total. Of these, 107 discussed
aspects of relativity and 58 put forth a concept of a velocity dependent mass. 
Similar results were obtained among surveys conducted in the San Francisco public
library, another key resource for the average citizen to access knowledge.

Another avenue for knowledge for the general public is television. Indeed, entire channels
have been devoted to the dissemination of science knowledge. No serious study is reported here but anecdotal evidence suggest that the concept of relativistic mass is still prevalent. 
In this International Year of Physics there have aired many programs
on Einstein's achievements both new and old. Two recent productions
display the typical scenario of explaining Noc via RM.
The first, airing in February 2005 on the Science Channel's special
``100 Greatest Discoveries''\cite{100greatest}, places relativity near
the top of their list and
in which Michio Kaku states that objects moving very fast, ``literally, grow heavier". 
In October, the two hour special ``Einstein's Big Idea"\cite{ebi} (based upon the book ``$E=mc^2$"
by David Bodanis\cite{bodanis})
explored the history of the concepts surrounding this equation in a well produced,
semi-dramatic documentary. Though not offering too many scientific details, Bodanis
does offer the same explanation as Kaku in regards to a train accelerating up to 
the speed of light, ``So all this energy, where does it go? It has to go somewhere.
Amazingly, it goes into the object's mass. From our point of view, the train actually gets
heavier."

Lastly, the most fashionable font of knowledge today is the world wide web, and within this realm
the starting point is often the Google search engine\cite{google}. 
A search for the term `special relativity' returned approximately 739,000 hits (the simpler
search for `relativity' primarily returned sites pertaining to general relativity). Now, the average person
will not stray too far from the first page, consisting of 10 links; thus, an exhaustive survey 
to gauge exposure to the concept can effectively be replaced by an examination of sites listed on the
first two pages. Of the first 20 links returned only two were not pedagogical expositions of 
relativity, and of the remainder 6 included a discussion of relativistic mass.

These anecdotal surveys support the claims put forth in this section; the concept
of relativistic mass is still widely embraced. Those ordinary citizens who strive to learn the theory
through self-study are likely to run across statements that relativity tells us that mass
increases with velocity.

\section{Conclusion}

	The modern theory of relativity relies upon the geometrical properties of spacetime as its foundation. The simplicity and beauty of the theory are regarded as its hallmark. A clear, precise definition of mass in this formalism arises naturally when defining momentum as the product of mass and 4-velocity. All primitive concepts are Lorentz covariant. The nature of mass is no longer simply Newtonian mass, as insisted by some\cite{okun}, but has a clear conception as expressed by Einstein\cite{einstein2}. 
By introducing a concept that is inferred and not primitive, that destroys the Lorentz covariance of the theory, 
relativistic mass is in direct conflict with the kinematical structure of special relativity. As the concept can in no way be considered a primitive concept of the theory, the statement that it is merely a matter of choice whether to use it or not is flagrantly incorrect.

For those who insist on continuing to use the concept of relativistic mass, serious reflection and examination are required. Is it possible to have a completely consistent velocity-dependent mass integrated into the full theory of relativity? Here we have seen it can not. Thus, is it wise to use it even as a heuristic device, as it may lead to erroneous conceptualization? Later studies will require
students to unlearn the concept as it is rarely employed in formal treatments.
One must justify its use with research that demonstrates its superiority over the geometric formulation.
Preliminary evidence introduced here suggests that it is not and furthers misconceptions.  

A survey of works that put forth a view of relativity show
that, for those industrious citizens that wish to tackle the theory 
of relativity on their own, an increasing majority of books to help in this endeavor profess
a mass that increases with velocity. However those that enroll in an institution of higher learning 
will increasingly be presented with a view not requiring a radical reconceptualization of mass.
A widening gap in the fundamental underlying meaning of relativity is afoot.

It is the view of this author that the continued introduction of such a problematic concept to the general public presents a consistency problem that must be addressed.
In fact, I find myself in agreement with Lev Okun\cite{okun2}, who states that this entails an ethical problem,
``Teaching the reader this formula usually entails deceiving him."
There have been several past calls to abandon this concept, however they have met
with only limited success. With the continual flood of books on the market that utilize this concept, many by non-experts,  more vigilance is required by the physics community in writing, reviewing,
and recommending such works.
In the centenary year of special relativity, we owe it to Einstein to get it right.

\begin{acknowledgments}
 I am indebted to Charles DeLeone for many helpful suggestions on the various drafts. I would also like to thank the Education Program for Gifted Youth and Patrick Suppes for support. 
\end{acknowledgments}


\begin{thebibliography}{1}
\bibitem{survey} The survey was given to the following groups of students:
Two groups (99 students in all)
consisted of general education science students at the college level. I am indebted to Charles DeLeone of C.S.U., San Marcos for his assistance. The third consisted of 58 gifted high school students (12
of whom were from Singapore). 
A fourth group consisted of 7 adults just beginning a basic modern physics course. None of these students had a formal introduction to special relativity at the time the survey was conducted.
\end{thebibliography}
\end{document}